\documentclass{elsarticle}

\usepackage[ruled,noline,slide, boxed]{algorithm2e}

\journal{Information Processing Letters}

\begin{document}

\begin{frontmatter}


\title{2-Dimensional Palindromes with $k$ Mismatches}


\author[mymainaddress]{Dina Sokol \corref{mycorrespondingauthor}}
\cortext[mycorrespondingauthor]{Corresponding author}
\ead[url]{http://www.sci.brooklyn.cuny.edu/~sokol}

\address[mymainaddress]{Department of Computer and Information Science, Brooklyn College and The Graduate Center, City University of New York,
 2900 Bedford Avenue, Brooklyn, NY 11210}

\newtheorem{observation}{Observation}


\begin{abstract}
This paper extends the problem of 2-dimensional palindrome search into the area of approximate matching. 
Using the Hamming distance as the measure, we search for 2D palindromes that allow up to $k$ mismatches.
We consider two different definitions of 2D palindromes and describe efficient algorithms for both of them.
The first definition implies a square, while the second definition (also known as a \emph{centrosymmetric factor}), can be any rectangular shape. 
Given a text of size $n \times m$, the time complexity of the first algorithm is $O(nm (\log m + \log n + k))$ and for the second algorithm it is $O(nm(\log m + k) + occ)$ where $occ$ is the size of the output.
\end{abstract}


\end{frontmatter}

\section{Introduction}

A \emph{palindrome} is defined as a string that reads the same backward as forward. When considering a multi-dimensional text, there are various ways to extend the palindromic properties.
In this paper we address two different kinds of 2-dimensional (2D) palindromes, each having different symmetry requirements. 2D palindromes have been given a lot of attention in  recent literature \cite{kulkarni2017two,mahalingam2018palindromic,mahalingam2019maximum,GEIZHALS2019161}. 

\emph{Approximate palindromes} are palindromes that admit errors of some kind.
In 1-dimension (1D), approximate palindromes have been studied from varying perspectives. When mismatches are allowed in the palindrome, 
they can be found in both run-length compressed texts \cite{chen_efficient_2012} as well as in the online model \cite{amir_approximate_2014}. Palindromes that admit insertions and deletions as well as mismatches (i.e. edit distance) are explored in \cite{hsu_finding_2009}. 

In this paper, we apply the concept of an \emph{approximate} palindrome to 2D. We allow up to $k$ mismatches in a palindrome, where $k$ is a given integer such that $k \geq 0$. We present algorithms that find all occurrences of two different kinds of 2D palindromes in a given 2D array, while allowing up to $k$ mismatches.

\begin{figure}
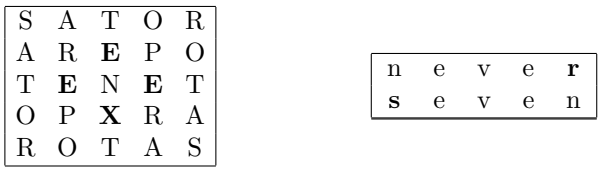

	\begin{minipage}[a]{0.4\linewidth}
		\centering
		\renewcommand{\tabcolsep}{4pt}
		\begin{tabular}{ | c c c c c | }
			\hline
			S & A & T & O & R \\
			A & R & \bf{E} & P & O \\
			T & \bf{E} & N & \bf{E} & T \\
			O & P & \bf{X} & R & A \\
			R & O & T & A & S \\
			\hline
		\end{tabular}
	\end{minipage}
	\begin{minipage}{0.4\linewidth}
		\centering
		\begin{tabular}{ | c c c c c | }
			\hline
			n & e & v & e & \bf{r}  \\
		      \bf{s} & e & v & e & n \\ \hline

           \hline
		\end{tabular}
	\end{minipage}

\caption {Depicted on the left is a  \emph{sq2DP} with one mismatch (character X), and on the right is a \emph{rect2DP}  with a single mismatch.}
\label{fig:2DPexamples}
\end{figure}

We recall the definitions of two kinds of 2D palindromes \cite{GEIZHALS2019161}. The  \emph{sq2DP} is an $m \times m$ 2D array that is symmetric over both diagonals (which implies a $180^\circ$ rotation as well). The \emph{rect2DP}  is an $m_1 \times m_2$ 2D array that is identical to its  $180^\circ$ rotation. 
Each 2D palindrome (of either type) has a \emph{center}, which is the point that results in an equal number of columns to the left and right, as well as an equal number of rows above and below. 
For odd dimensions, the center is at a character in the text, while for even dimensions the center can land between characters.
For example, the center of the \emph{rect2DP} in Figure~\ref{fig:2DPexamples} is the third column, between the two v's.

In a 1D palindrome, the meaning of a mismatch is a pair of 2 mismatching characters, for example, the string \emph{abccXa} is a palindrome with one mismatch between the \emph{b} and \emph{X}. In \emph{rect2DP} this definition extends trivially since each character must match exactly one other character. However, in \emph{sq2DP}, it is necessary to define what a mismatch means, as several symmetries are involved.

We call a \emph{layer} of a \emph{sq2DP}, one square consisting of 2 rows and 2 columns that are equi-distant from the center. For example, 
in Figure~\ref{fig:2DPexamples} the outermost layer consisting of SATOR four times would be considered layer 2. 

\begin{observation}
Each character in a \emph{sq2DP}, except those on a diagonal, must be equivalent to three other characters in its layer. 
\end{observation}

This observation results in equivalence classes, each containing exactly four locations. To count mismatches in a \emph{sq2DP} we take the majority character in each equivalence class. The characters not equal to the majority are considered mismatches. In case of a tie, any winning character is chosen arbitrarily.
For example, in Figure~\ref{fig:2DPexamples}, the character X in layer 1 is the mismatch, since the majority in its equivalence class is E. This \emph{sq2DP} is said to contain 1 mismatch.

A \emph{$k$-mismatch} 2D palindrome (of any type) is a 2D palindrome that contains no more than $k$ mismatches.
The concept of maximality applies to 2D palindromes with mismatches, where a $k$-mismatch palindrome is \emph{maximal} if extending in any direction, maintaining the same center, results in a palindrome with more than $k$ mismatches.

In this paper we solve the  
2D palindrome with $k$-mismatches problem. Given a 2D text $T$ of size $n \times m$, the goal is to find all maximal 2D palindromes with $\leq k$ mismatches.
We present two different algorithms, one for each definition of a 2D palindrome.
In Section~\ref{sec:sq2DP} we describe an algorithm to search for $k$-mismatch \emph{sq2DP}, which have multiple symmetries, yet we are able to solve this problem in $O(nm (\log m + \log n + k))$ time. In Section~\ref{sec:rect2DP} we obtain a reduction of the problem of locating all maximal \emph{rect2DP} to the 1D problem, resulting in a $O(nm(\log m + k) + occ)$ time algorithm, where $occ$ is the size of the output. 



\section{Approximate Square 2D Palindromes}
\label{sec:sq2DP}

{\bf \noindent Problem Definition: } Given an $n \times m$ 2-dimensional text $T$, find all maximal $k$-mismatch \emph{sq2DP} in $T$.

\subsection{Straightforward Algorithm}

We begin by presenting an  algorithm to find all \emph{sq2DP} that is based upon known techniques and is straightforward, yet has cubic time complexity. 
First, the text $T$ is preprocessed by constructing a generalized suffix tree (GST) \cite{ukkonen_-line_1995} for the columns of $T$, from bottom to top and from top to bottom, and for the rows of $T$, from left to right and right to left. Then, the GST is preprocessed for lowest common ancestor queries (LCA) \cite{harel_fast_1984} to allow $O (1)$-time longest common prefix (LCP) queries \cite{LV-89,LV:88}.


The algorithm considers each possible center (both at characters and between characters) and tries to find the maximal \emph{sq2DP} with no more than $k$ mismatches at that center. Note that since this kind of palindrome is square,
there is at most one maximal palindrome occurring at each center. The idea is to grow the palindrome from the center, by matching corresponding row/col pairs in each layer. 
Consider a layer that we are attempting to add on to an existing \emph{sq2DP}; see Figure~\ref{fig:trapNum} for an illustration. In order to add a layer, comparisons must be done between each side and its reflection over both the main diagonal and the anti-diagonal. 
Four LCP queries suffice to accomplish this.

The worst case time complexity of this algorithm is $O(nm^2)$. Consider the input text that consists of all a's. The suffix tree construction and LCA preprocessing can be done in linear time \cite{ukkonen_-line_1995,harel_fast_1984}. Each possible center, of which there are $O(nm)$, will do $O(min(n,m))$ LCP queries in constant time, yielding $O(nm^2)$ time overall (w.l.o.g. we can assume $m \leq n$).

It is easy to extend this algorithm to search for $k$-mismatch \emph{sq2DP}.
Given a center, $(r,c)$, once again attempt to add one layer at a time to the palindrome. 
If the corresponding sides match fully, then the layer is added,
and the algorithm continues with the next layer. If there is a mismatch in a layer, then LCP's must be done to locate all mismatches within the layer. For each mismatch found, the equivalence class of locations is checked, and the number
of mismatches is counted. A tally is kept, and the algorithm finishes with this center either when the outer edge of the text is reached, or $k$ mismatches are found.
The worst case time complexity of the algorithm that allows $k$ mismatches is also $O(nm^2)$  (assuming $k < m$, otherwise it is $O(nmk)$).

\subsection{Improved Algorithm for Approximate Sq2DP}

The goal of the improved algorithm is to check the sides of several layers at once, attempting to locate the first $k$ positions of mismatch. 
The difficulty in checking neighboring sides together is that the length of a side grows by two in each subsequent layer. However, this yields a trapezoid shape as shown in  Figure~\ref{fig:trapNum}. Thus, we begin by solving the problem of matching two trapezoids.


\subsubsection{Matching Trapezoids}
\label{sec:trapezoids}

Given two 2D texts, each of the same trapezoid shape, with height $n$, and width of widest row $m$, the goal is to match them to find the longest prefix of the trapezoids, from top to bottom, that matches with $\leq k$ mismatches, and to locate the locations of
mismatch. If we preprocess the trapezoids by constructing suffix trees of their rows, we can answer the query in time $O(min(n,k))$. 
However, if we would incorporate this into the algorithm for \emph{sq2DP}, it would result in cubic time complexity. In order to reduce the overall time, we present an algorithm that does a little extra initial processing, and then answers the query more efficiently.
Thus, we apply a naming technique to the rows of the trapezoids, as follows. For each row, we name its prefix and suffix of length $2^i$, such that
$2^i$ is the largest power of 2 that is smaller than the row's length. 
In essence, we are creating $O(\log m)$ strings of names, such that each string of names contains the names for a particular $2^i$, and these strings lie on the left and right diagonal edges of the trapezoid. We then preprocess the strings of names that lie
 on the left and right edges of the trapezoid to allow LCP queries.

When given two trapezoids to match, we will do $O(\log n)$ LCP queries between diagonals as follows. In the spirit of KMR \cite{KMR}, there is some $i$ such that the width of the first row of the trapezoid $w$, satisfies
$2^i \leq w < 2^{i+1}$.
We perform LCP queries on the diagonals of names, beginning with the names of width $2^i$. These queries will have to be done on both of the outer diagonals of the trapezoids, since the width $2^i$ provides
overlapping prefixes and suffixes of each row. In the event that a mismatch is found, its row is compared to locate the exact positions of mismatch, using the regular suffix tree of the rows.
The $2^i$ names work though only while the width of the trapezoid row is $< 2^{i+1}$. This is where the $\log n$ queries come in to play. For each set of rows of width $2^j \leq w' < 2^{j+1}$, another LCP query must be done, 
with the names of width $2^j$. For example, if the width of the first row of the trapezoid is 4, the diagonal of names of width 4 will be used to match the rows that have width 4-7. Beginning in the row with width 8, the names
of width 8 will be used to match rows with width 8-15, and so on.

\begin {algorithm} 
\LinesNumbered
\caption{Matching Trapezoids}
\label{alg:trapezoids}
\SetKwInOut{Input}{Input}
\SetKwInOut{Output}{Output}
\SetKwArray{Kw}{$T$}

\Input {Two trapezoids, width of first row is $w$, bottom row is $m$. GST of the rows of the trapezoids, and GST of the outer diagonals of names of widths $2^j$ such that $w \leq 2^j \leq m$.}
\Output {Locations of first $k$ mismatches between the two trapezoids.}
\BlankLine
$i = \lfloor \log w \rfloor$ \\

$k' = 0$  \tcc*[f]{$k'$ keeps track of the number of mismatches} \\
\While { $k' < k$ AND $2^i < m$ }
   {
      use diagonals of names with width $2^i$ to compare rows with width $2^{i}$ to width $2^{i+1}-1$

      for each mismatch between names, use suffix tree of rows to locate actual mismatches 

      increment $k'$ each time a mismatch is found

      $i++$ \tcc*[f]{increment $i$ for next set of rows}
    }   

\end {algorithm}

The time complexity for this algorithm is as follows. 
Assume $m$ is the maximum width of the trapezoids, and $n$ is the height. Constructing the $\log m$ texts of names can be done in time linear to the size of the trapezoid for each $2^i$, or $O(nm \log m)$.  
Then, constructing the suffix trees of diagonals in each of these texts can be done in the same amount of time. 
(Technically we can get this down to linear in the size of the trapezoid since each row needs only 2 names. However, the improvement does not help in the larger scheme of the algorithm for \emph{sq2DP}.)
The comparisons will do at most $\log n + k$ LCP queries, and hence will take
$O($max($k, \log n$)) time. Overall, this results in $O(nm \log m)$ for the initial processing and data structure construction, and then $O($max($k, \log n$)) query time, for a total of  $O(nm \log m + \log n + k)$ time. 

\subsubsection{Converting layers to trapezoids}

The idea of the improved algorithm is to do the preprocessing of Algorithm~\ref{alg:trapezoids} once for the entire text, naming each subrow of
 length $2^i$, for $2 \leq i \leq \log m$.
A GST is constructed, for each text of names, of the forward diagonals and anti-diagonals. 
Forward diagonal $d$ for $1 \leq d < m+n$ is defined as all locations $(r, d-r)$ such that $1 \leq r \leq min(n,d)$. 
Anti-diagonals are defined analogously.

\begin{figure}
	\centering
	\includegraphics [width=200pt] {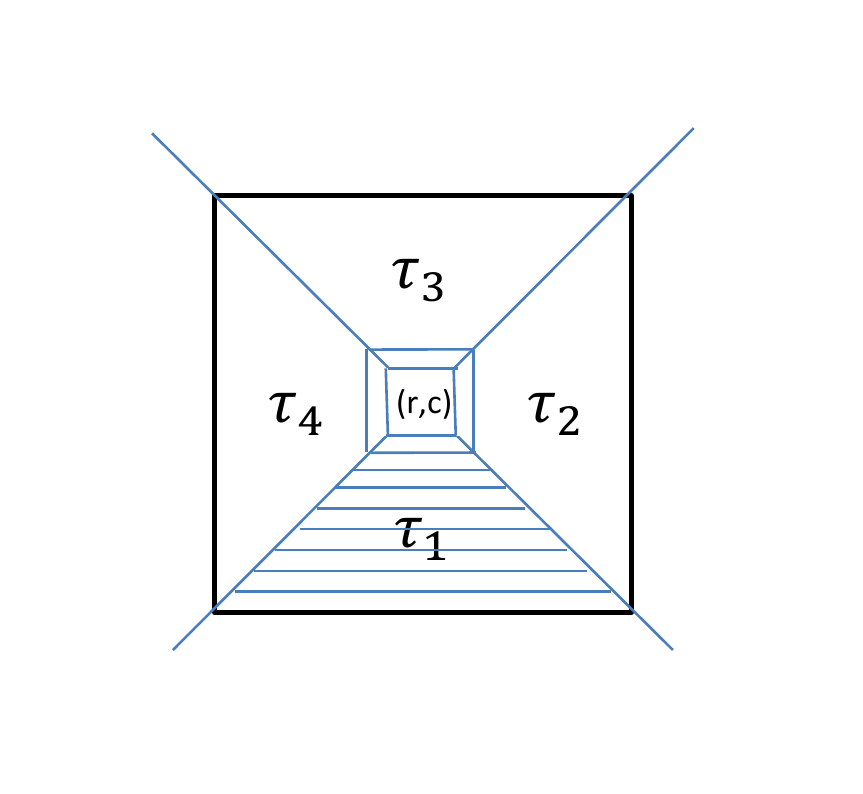}
\caption {The location $(r,c)$ is the current center. A layer around the center is shown. Several layers can be depicted by a set of four trapezoids, numbered $\tau_1$-$\tau_4$.} 
\label{fig:trapNum}
\end{figure}

We then check each center by matching trapezoids around that center.
See Figure~\ref{fig:trapNum} that depicts several contiguous layers and marks how they turn into trapezoids. In the improved algorithm, instead of extending by individual layers, we match the trapezoids in each direction,
finding the locations of the first $k$ mismatches. Since there are two directions of symmetry, as in the previous algorithm, each trapezoid has to be matched twice. But each time a mismatch is found, its equivalence class is processed,
and we make sure to count the mismatches in each equivalence class once. See pseudocode in Algorithm~\ref{alg:sq2DPTrapezoids}.

\begin {algorithm} 
\LinesNumbered
\caption{$k$-mismatch \emph{sq2DP}}
\label{alg:sq2DPTrapezoids}
\SetKwInOut{Input}{Input}
\SetKwInOut{Output}{Output}
\SetKwArray{Kw}{$T$}
\Input {Text $T$ of size $n \times m$, integer $k$.} 
\Output {All maximal \emph{sq2DP} with  $\leq k$ mismatches.}
\BlankLine

Construct a GST preprocessed for LCA of all rows of $T$, both forward and reverse. \\

Do naming of subrows of widths (a power of 2) $2^i$, for $1 \leq i \leq \log m$.\\

Construct GST preprocessed for LCA of all diagonals of names of each size, computed in the previous step. 

\For {each center $(r,c)$}

Number the trapezoids around location $(r,c)$ with $\tau_1$-$\tau_4$, counterclockwise, as shown in Figure~\ref{fig:trapNum}.
\\
Call Algorithm~\ref{alg:trapezoids} on the following trapezoid pairs: $(\tau_1,\tau_2), (\tau_4,\tau_3), (\tau_2,\tau_3), (\tau_1,\tau_4)$. \\

Merge the errors found in the previous step, from inside to outside, stopping when $k$ is reached. \\

Output the distance of the final layer added. 
\end {algorithm}

Time Complexity of improved algorithm: Given an $n \times m$ text, we preprocess the entire text 
by naming every subrow of width $2^i$, creating $\log m$ texts of names, in $O(nm \log m)$ time. The diagonals
are preprocessed in both directions to be placed in a GST preprocessed for LCP queries in the same amount of time. Each center is checked by calling Algorithm~\ref{alg:trapezoids} four times with time complexity $O(\log n + k)$.
The merge takes no more than $O(k)$ time. Thus, the overall time for the algorithm is $O(nm (\log m + \log n +k))$.

\section{Approximate Rectangle 2D Palindromes}
\label{sec:rect2DP}

The main difference between \emph{rect2DP} and \emph{sq2DP} is that \emph{rect2DP} have only one symmetry, the $180^\circ$ rotation, while  \emph{sq2DP} have both diagonal
symmetries as well. One implication of this is reflected in their names, as \emph{rect2DP} can be any rectangular shape, while \emph{sq2DP} must have size $x \times x$.
On the one hand, the problem of \emph{rect2DP} seems simpler due the single symmetry, yet the difficulty of this problem is that each center may have $O(m)$ \emph{rect2DP} since each width can result in a maximal palindrome of a different height. With \emph{sq2DP}, there is exactly one maximal output for a given center. For exact \emph{sq2DP}, the algorithm of \cite{GEIZHALS2019161} takes advantage of this to yield a linear time algorithm for \emph{sq2DP}. For \emph{rect2DP}, two algorithms are given with a tradeoff in \cite{GEIZHALS2019161}, the first has time complexity $O (nm^2)$, 
and the second has time $O(nm \log n + occ \log n)$. The second algorithm spends $O(\log n)$ time per occurrence found, and is more efficient if the output size is small. However, for the worst case output size of $O(nm^2)$, the first algorithm would prove to be more efficient.
In this paper, we realize that similar techniques can be applied in a more clever way to yield both a more efficient algorithm for all cases, and one that extends naturally to solve the $k$-mismatch problem as well. 
\\

{\bf \noindent Problem Definition: } Given an $n \times m$ 2-dimensional text $T$, find all maximal $k$-mismatch \emph{rect2DP}  in $T$.
\\

Given a 1D text, a well-known algorithm to solve the problem of finding all palindromes works as follows. Construct a generalized suffix tree (GST) of both the text and its reverse, and preprocess the GST for  lowest common ancestor (LCA) queries that allow constant time Longest Common Prefix (LCP) queries. For each possible center, do an LCP query in the forward and reverse text, yielding the output for this center.  For finding all $k$-mismatch palindromes, $k+1$ LCP queries are done for each center, sometimes referred to informally as ``kangaroo jumps'' \cite{GG-86}.

\begin{observation}
A 2D array is a \emph{rect2DP} if and only if it has an exact match between its lower half and the $180^\circ$ rotation of its upper half.
\end{observation}

\begin{figure}
	\centering
	\includegraphics [width=120pt] {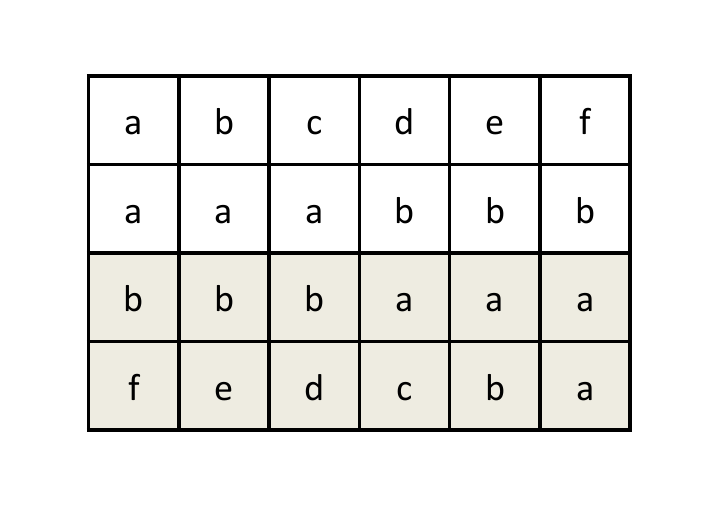}
      \caption {Comparing a $180^\circ$ rotation of the array to itself is identical to comparing its bottom half with a $180^\circ$ rotation of its top half. For odd height, include the middle row in \emph{both} halves.}

      \label{fig:rect2DP}
\end{figure}

If you rotate a \emph{rect2DP} and match it to itself, then you will be doing each comparison twice. Hence, matching halves suffices; see Figure~\ref{fig:rect2DP} for an example. This observation yields a simple reduction from the 2D problem to the 1D problem, \emph{for a given width}. Preprocess the 2D text so that each subrow is given a name. This can be done during the construction of a GST of all rows of the text. 
Then, preprocess the columns of names for LCP queries.

When processing width $w$, simply run the 1D algorithm on the columns of names of width $w$. For each center, perform an LCP query upward and downward in its column of names. For the $k$-mismatch case, do $k+1$ LCP queries (as in 1D), but for each mismatch between names, compare the rows that differ and tally the actual mismatches, up to $k$. The same GST that is used to name the subrows can be used to compare the names that differ. No more than $O(k)$ work is done per center, per width. Hence, the time complexity for this algorithm is $O(n m^2 k)$, as both the naming and the searching can be done in this time. If $k=0$, the time is  $O(nm^2)$. 

\begin{figure}
	\centering
	\includegraphics [width=250pt] {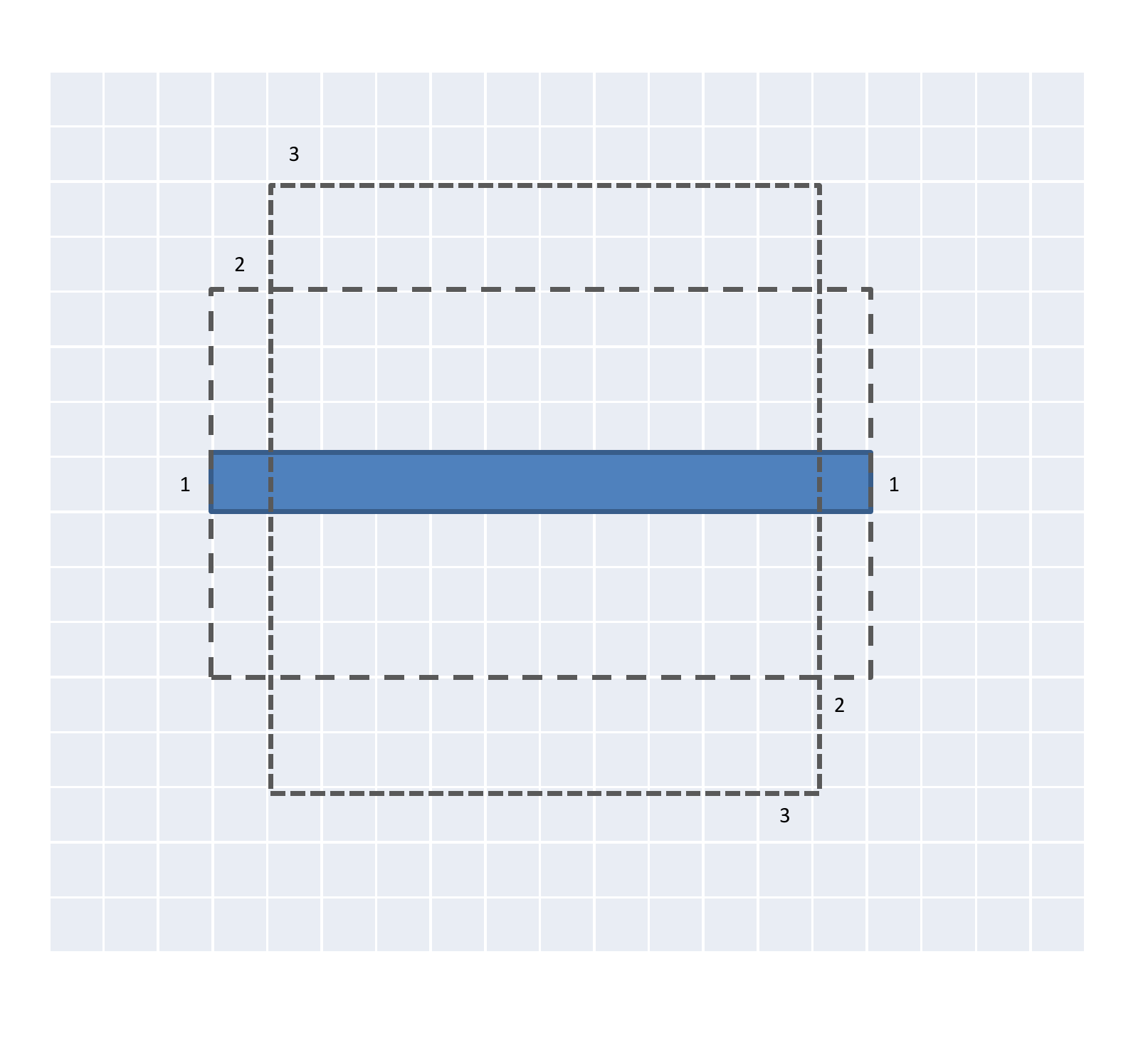}
      \caption {The two rectangles with dotted lines are maximal \emph{rect2DP} for the given center (with no mismatches). Positions of mismatch are denoted with numbers, in the order in which they are located.}
      \label{fig:noMismatch}
\end{figure}

We now apply the idea that it is only desirable to do work when there is output. We first improve the algorithm without mismatches, and then generalize for $k$ mismatches.
As observed in \cite{GEIZHALS2019161}, if there is no output for a given width, then it is not necessary to check any larger widths. Hence, we can begin with the
 widest possible width and then figure out which smaller widths are necessary to check as follows. Begin by finding height 1 (and 2 for even heights) using the 1D algorithm (for height 2, do 2 suffix/prefix LCP queries). Whatever width is returned by this initial iteration is checked in constant time, by doing an LCP query on the names of width $w$, as in the algorithm described in the previous paragraph. 
The position of mismatch tells the row in which the palindrome must get thinner. This row is checked trivially, as was done in the first step of the algorithm.
If it is impossible to obtain output with the next row, then the center is finished being processed. 
Otherwise, that row tells the next width that needs to be checked. 
See Figure~\ref{fig:noMismatch} for an example. 
Note that the number of widths that are checked is $O(occ)$. The time complexity for this search is $O(nm + occ)$  since the number of centers is $O(nm)$, each is processed in $O(occ)$ time.

The only remaining issue is that if the naming is done naively, it takes cubic time to write down $O(nm)$ names of each width. However, we can use the improved naming technique of \cite{GEIZHALS2019161} to name only widths a power of 2. Each width that is not a power of 2 will have two overlapping names that are a power of 2, and hence can still be checked in constant time. This results in an overall time complexity of $O(nm \log m + occ)$.

\begin{figure}
	\centering
	\includegraphics [width=250pt] {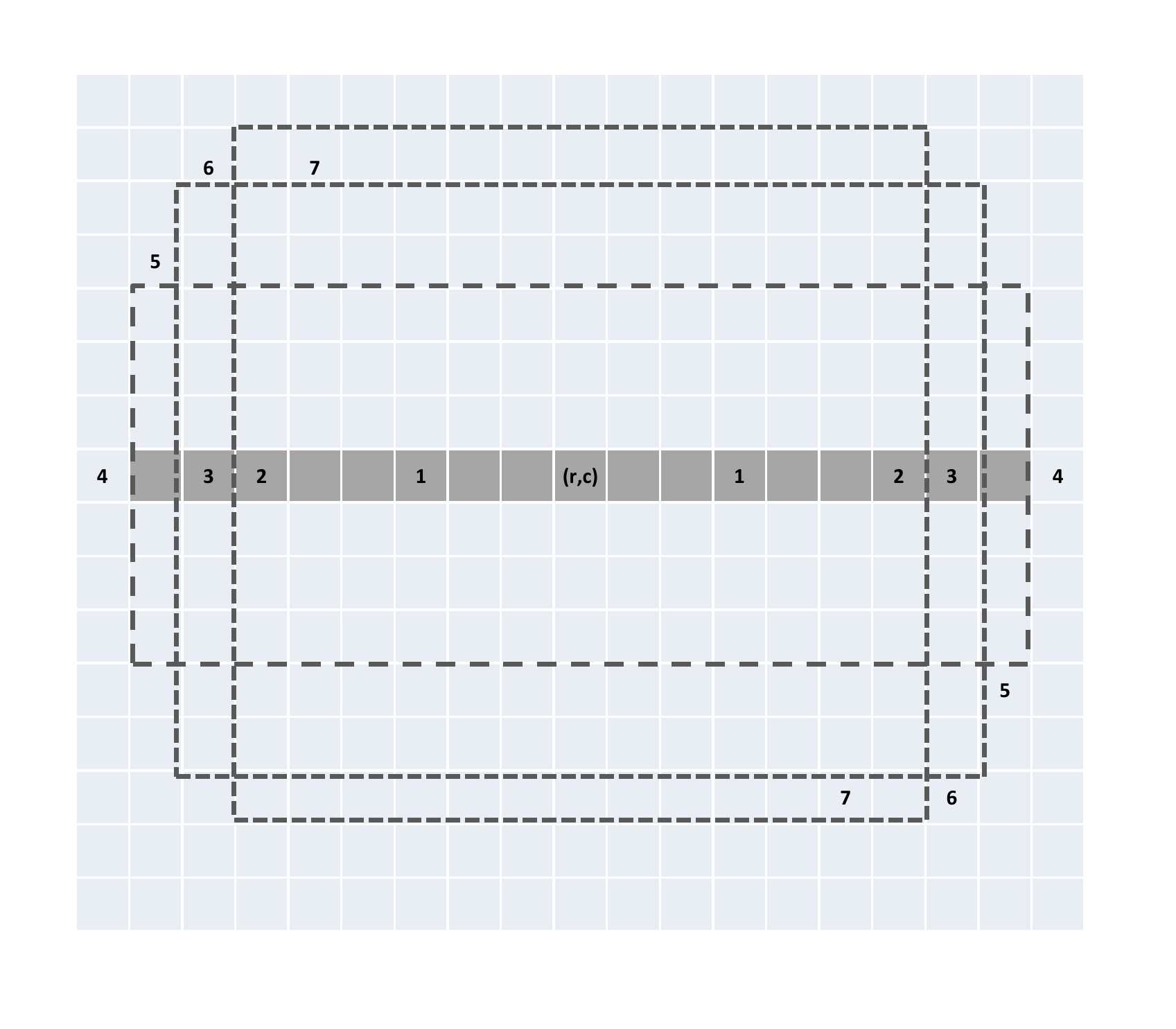}
      \caption {Shown are three maximal $k$-mismatch \emph{rect2DP} for a given center, with $k=3$ mismatches. Positions of mismatch are denoted with numbers, in the order in which they are located. The second output is found as a thinner width of 3 mismatches, while the third output is found by removing mismatch 3 and allowing for another mismatch in the palindrome lower down (mismatch 7).}
      \label{fig:kMism}
\end{figure}

Algorithm~\ref{alg:rect2DP} and Figure~\ref{fig:kMism} illustrate how this algorithm will extend to allow $k$ mismatches.
It begins by using the 1D algorithm with $k$ mismatches on the given center $(r,c)$.
With $k$ mismatches in the center row, we can only extend the entire width if we allow  zero additional mismatches. When we shrink the width, if we do not knock off any errors, we can still allow no mismatches in the extension (as in the algorithm for $k=0$). However, once we cannot extend any more while allowing no mismatches, it is necessary to knock off a mismatch from the edge, and allow an additional mismatch when going down. This is illustrated in Figure~\ref{fig:kMism} with mismatch number 7.

Time Complexity:
We still have the preprocessing time of $O(nm \log m)$ for the naming and GST constructions.
When processing a center, $O(k)$ work is done for the 1D algorithm on the center row. For the extensions, the first type of checking (getting skinnier with no new mismatches) can be charged to the output. The second type will be no more than $O(k + occ)$ work per center, since each mismatch is knocked off once for the non-extendable case, and if it can be extended we charge to the output. Hence, the overall time is $O(nm(\log m + k) + occ)$.
\begin {algorithm} 
\LinesNumbered
\caption{Algorithm for $k$-mismatch \emph{rect2DP}}
\label{alg:rect2DP}
\SetKwInOut{Input}{Input}\SetKwInOut{Output}{Output}
\SetKwArray{Kw}{$T$}
\Input {2D text $T$ of size $n \times m$,  GST of rows of $T$ in forward and reverse order, and the $\log m$ texts of names preprocessed columnwise, up/down, for LCP queries.}
\Output {All maximal $k$-mismatch \emph{rect2DP}(s) in $T$.}
\BlankLine
\For {each center $(r,c)$}
    Run the 1D algorithm and find the maximal $k$-mismatch palindrome of heights 1 and 2. Let $w$ be the width of the palindrome returned.
\\
    Use the texts of names appropriate for width $w$ to extend this output to its maximal height. Let $w'$ be the width of the palindrome
extended into the row beneath the maximal height found. 
\\
   Attempt to extend width $w'$. Keep repeating Lines 3 and 4 until no longer extendable with zero mismatches.
\\
  Knock off $k$th mismatch and attempt to extend down, allowing another mismatch. Keep knocking off mismatches from outermost edges and extending down as many times as possible.
\end{algorithm}

\newpage
\section*{References}


\bibliography{IPLpals.bib}

\end{document}